\documentclass[epj]{svjour}
\usepackage[abs]{overpic}
\usepackage{graphicx}
\usepackage{rotating}
\usepackage{amssymb,amsmath}
\usepackage[square,comma,sort&compress]{natbib}
\usepackage{citesort}
\usepackage[dvips]{color}
\begin{document}
%
\title{
On the nature of new baryon state X(2000) observed in the experiments
with the SPHINX spectrometer.
}
\author{
L.G.~Landsberg%
}                     
\institute{
State Scientific Center of Russian Federation\\
``Institute for High Energy Physics''\\
142280, Protvino, Moscow region, Russian Federation\\
E-mail: LeonidLandsberg@ihep.ru}
\authorrunning{L.G.~Landsberg}
\titlerunning{On the nature of new baryon state X(2000)}
%
%
\abstract{%
In the experiment with the SPHINX spectrometer in 1995-99 a new baryon state
$X(2000) \to \Sigma^0K^+$ was observed in the proton diffractive reactions
$p + N(C) \to \Sigma^0K^+ + N(C)$.
The main parameters of $X(2000)^+$ baryon are 
$M = 1986 \pm 6 MeV , \Gamma  = 98 \pm 21 MeV$.
Unusual features of massive X(2000) state 
(narrow decay width $\frac{\Gamma}{M} \simeq 0.05$, 
anomalously large branching ratio with strange particle emission) 
make it serious candidate for cryptoexotic pentaquark baryon 
with hidden strangeness.
New data on the SPHINX with by an order of magnitude enlarge statistics are 
in agreement with previous data and are supported our conclusions.
We propose further studies X(2000) baryon production in the meson reactions 
with baryon exchange: $\pi^{\pm} + p \to X(2000)^+ + \pi^{\pm}$
(OZI - forbidden suppressed reaction 
for the state $\vert qqqs \bar{s}\rangle$) and
$K^{\pm} + p \to X(2000)^+ + K^{\pm}$
(OZI - allowed reaction).
If the value of
$\mathrm{R}=\frac
{\mathrm{BR} \lbrack \pi^{\pm} + p \to X(2000)^+ + \pi^{\pm} \rbrack} 
{\mathrm{BR} \lbrack   K^{\pm} + p \to X(2000)^+ +   K^{\pm} \rbrack}
\ll 1$ 
it will be the crucial argument in favor of 
$X(2000)= \vert qqqs \bar{s}\rangle$) structure.
These experiments may be done on IHEP separated kaon beam 
 with OKA spectrometer.
} 
\maketitle
\section{Introduction}
\label{se-1} 

\begin{figure}
\vspace*{-2cm}
\includegraphics{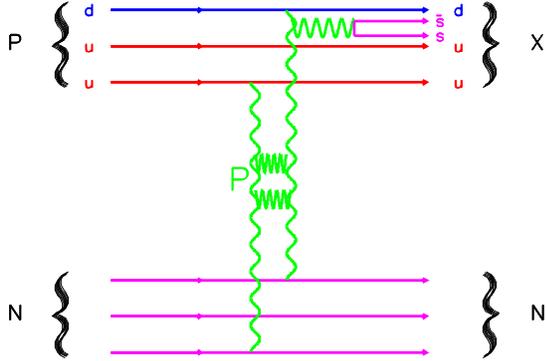}
\vspace*{-2cm}
\caption{%
Diagram for production of an exotic hadron with hidden strangeness %
$\vert X \rangle = \vert uuds \bar{s}\rangle$ in difractive processes %
due to Pomeron exchange. %
The main component of Pomeron P represents a gluon ``ladder''. %
In such gluon enriched reactions the probability of exotic hadrons %
maybe quite high.%
}
\label{fig-fnm-diff}
\end{figure}

In the experiments with SPHINX spectrometer an extensive program 
of studying diffractive production proton reactions on nucleons and 
nuclei (coherent processes) was carried out 
on the secondary proton beam of IHEP accelerator 
with energy $E_P = 70 GeV$ and 
intensity   $I_P =(2 \div 4) * 10^6 p/spill$ . 
The main aim of this program is the search for cryptoexotic baryons 
with hidden strangeness $\vert qqqs \bar{s}\rangle$ (here q=u- or d-quarks). 
This program was discussed in detail in 
the review papers~\cite{ref_01, ref_02, ref_03}. 
These searches were performed primarily in the gluon-enriched Pomeron
exchange diffractive production reactions. 
According to modern ideas, the main component of a Pomeron is sort
of gluon "ladder", which can provide for Pomeron processes a special role 
in production of exotic hadrons (Fig.~\ref{fig-fnm-diff}).
   Of significant interest are coherent diffractive processes
taking place on the nucleus as a whole. 
We shall deal with these processes with more detail and 
discuss method for identification of them. 
Consider a diffractive production of a certain set of secondary particles, 
for example $p + A \to \lbrack abc \rbrack + A$, which can proceed coherently 
on the nucleus $\mathrm{A}$. 
For identification of a coherent processes we consider the distribution 
of events for the reaction of interest over 
the square of transverse momentum $P^2_T$. 
In accordance with the uncertainty principle a coherent process 
proceeding on a nucleus as a whole is characterized 
by relatively small transverse momenta $P_T$ inversely proportional 
to the radius of the target nucleus $R = cons \cdot  A^{\frac{1}{3}}$ . 
A coherent process manifests itself as a narrow diffractive peak 
in the distribution of events over $P^2_T$ (see below).

\begin{figure}
\includegraphics{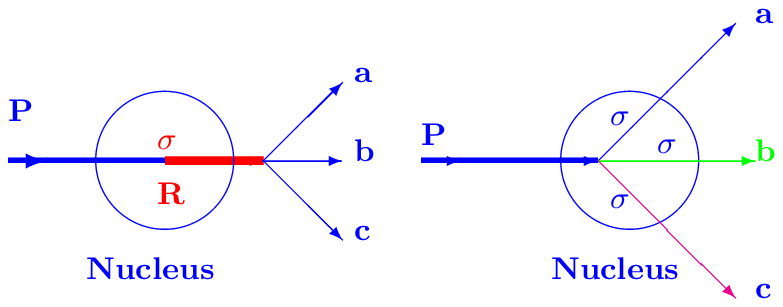}
\caption{%
Schematic illustration of suppression of the non-resonance multiparticle %
background with respect to the production of a resonance in the coherent %
reactions (the so-called coherent filter). %
\newline
$p + \mbox{(nucleus)} \to R + \mbox{(nucleus)}, R \to a + b + c$
\newline
(reaction involving production and cascade decay of resonance R), and %
\newline
$p + \mbox{(nucleus)} \to \lbrack a + b + c \rbrack + \mbox{(nucleus)}$
\newline
(non-resonance multiparticle background). Multiparticle states %
interact in the nucleus with a higher probability and violate %
the condition of coherence more strongly than in case of a resonance %
$\mathrm{R}$ ($\sigma $ represent the interaction cross section of %
secondary particles inside the nucleus). Therefore, the following %
relationship holds valid for the cross section of the processes at issue : %
\newline
$\frac{\sigma_{\rm coh}\mbox{(reson.)}}
      {\sigma_{\rm coh}\mbox{(nonres.backgr.)}} > %
 \frac{\sigma_{\rm noncoh}\mbox{(reson.)}}
      {\sigma_{\rm noncoh}\mbox{(nonres.backgr.)}} %
$ %
\newline
Coherent processes on nuclei act like a filter 
in making the identification of hadron resonances more unambiguous.%
}
\label{fig-filter}
\end{figure}

    Coherent processes serve as a certain filter that permits more clear 
identification of the produced resonances $R \to a+ b + c$ 
with respect of non-resonance multiparticle background. 
In the case of multiparticle events the probability of secondary interactions
in the nucleus exceeds the respective probability for resonances. 
Secondary interactions violate the condition of coherence. 
Therefore, in the case of coherent events the non-resonance background can
be significantly reduced with respect to the resonance effects. 
These arguments are qualitatively illustrated 
by the diagrams depicted in Fig.~\ref{fig-filter}.

   The experiment for studying diffractive production reaction 
were carried out on the SPHINX spectrometer. 
This multipurpose facility includes wide aperture magnetic spectrometer 
with proportional chambers, drift tubes and hodoscopes, hodoscopical 
electromagnetic spectrometer with lead-glass counters, 
Cherenkov detectors for identification of secondary particles, 
the guard system for separation of exclusive diffractive-like reactions. 
In the process of measurements the SPHINX spectrometer has gone through 
several modifications. 
All measurement with this setup can be divided into two stages:\\
a) The experiments of the first generation with "old" SPHINX 
   (in runs of 1989 - 1995). 
   The main results of these measurements were published in 1994 - 2000 
(see ref.~\cite{ref_04, ref_05, ref_06, 
                ref_07, ref_08, ref_09, ref_10, ref_11, ref_12, 
                ref_13, ref_14, ref_15, ref_16, ref_17, ref_18}
and review papers~\cite{ref_01, ref_02, ref_03}). 
   Short description of "old" SPHINX are presented in ~\cite{ref_04, ref_14}.\\
b) The experiments of the second generation with completely upgraded SPHINX 
   detector (in runs of 1996 - 1999).

   "Old" and completely upgraded SPHINX spectrometer had 
the same general structure. 
But after upgrade the facility was equipped with a new tracking system, 
new hodoscopes, hadron calorimeter, new electronics, DAQ and 
on-line computers (which increase the maximal available flux of data per spill
more than by order of magnitude). 
As a result of the upgrade, we have obtained a practically new setup, 
which is described in ~\cite{ref_19, ref_20}.

   With the upgraded setup during the runs 1996 - 1999 more than 
$10^9$ events were recorded on magnetic tapes.
This statistics now used to study different physical processes. 
First result of these studies are published in ~\cite{ref_20, ref_21, ref_22}.


\section{The main result for X(2000) baryon obtained in the first 
         generation of experiments with the SPHINX detector.}
\label{se-2}

   The most important result obtained in the first generation of experiments 
with the SPHINX spectrometer is observation of new X(2000) baryon 
in the reaction
\begin{eqnarray}
\label{re-01}
p + N(C) &\to &X(2000) + N(C)                                  \\
         &    &\phantom{[}\raisebox{.4cm}{\rotatebox{180}{$\Lsh$}}
               \Sigma^0K^+;\Sigma^0 \to \Lambda\gamma; \nonumber
\end{eqnarray}
This result was obtained in studying the reaction
\begin{eqnarray}
\label{re-02}
p + N (C) \to \lbrack \Sigma^0K^+ \rbrack + N (C)
\end{eqnarray}
(see ref.~\cite{ref_01, ref_02, ref_03, ref_04, ref_05, ref_06, 
                ref_07, ref_08, ref_09, ref_13, ref_14, ref_17}\footnote{
N(C) means that reaction is on quasi-free nucleons N or in coherent process 
on carbon nuclei as a whole}.
\begin{figure}
\includegraphics[width=\hsize]{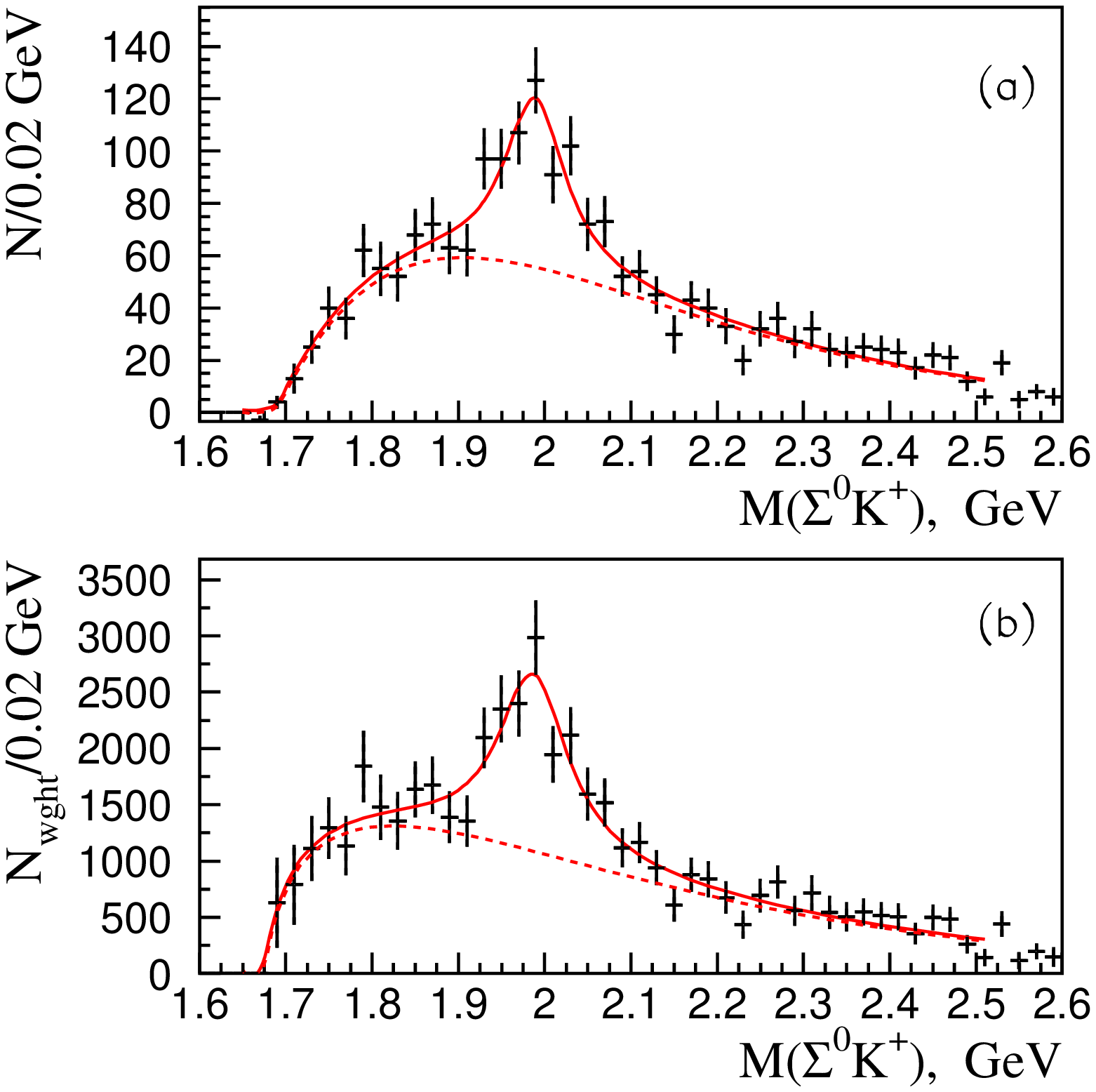}
\caption{%
Effective mass spectrum $M(\Sigma^0K^+)$ in the diffractive reaction %
$ p + N \to  \lbrack \Sigma^0K^+ \rbrack +N $ %
for the entire range of transverse momentum $P^2_T$ : %
\newline
(a) raw data; (b) spectrum weighted with the efficiency of the setup. %
A clear peak with parameters $M=1986\pm6MeV, \Gamma=98\pm21MeV$ is seen %
in the spectrum owing to production of X(2000) baryon.
}%
\label{fig-3}
\end{figure}
   Fig.~\ref{fig-3} presents the effective mass spectrum 
in~(\ref{re-02}) for the entire range of transverse momentum $P^2_T$. 
A clear peak with parameters
\begin{equation}
\label{re-03}
M = 1986 \pm 6 MeV; \Gamma  = 98 \pm 21 MeV 
\end{equation}
is seen in this spectrum and thus reaction~(\ref{re-01})
on quasi-free nucleons N was identified. 
\begin{figure}
\includegraphics[width=\hsize]{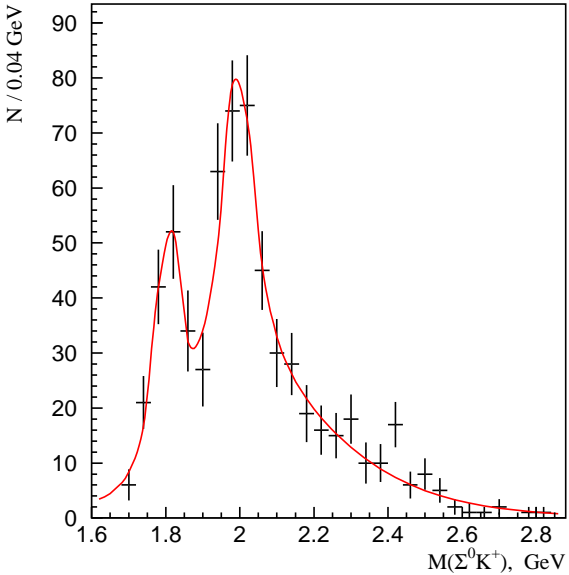}
\caption{%
Effective mass spectrum $M(\Sigma^0K^+)$ in the coherent diffractive reaction %
$p + C \to \lbrack \Sigma^0K^+  \rbrack + C$ %
with small transverse momenta $P^2_T < 0.1 (GeV/c)^2$ (coherence condition). %
Besides the near-the-threshold structure of X(1810) with mass %
$M \simeq 1807$ MeV , the spectrum has a dominant clearly defined peak %
of the X(2000) state.
}%
\label{fig-4}
\end{figure}
\begin{figure}
\includegraphics[width=\hsize]{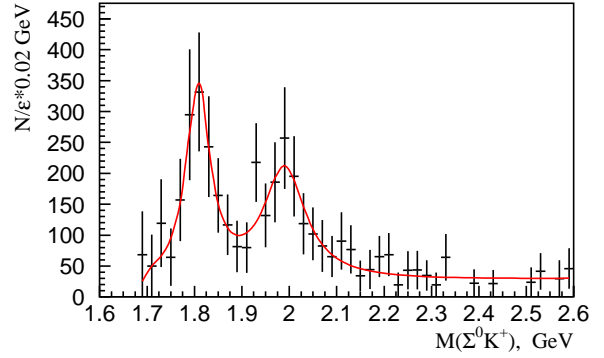}
\caption{%
Effective mass spectrum $M()$ in the coherent reaction %
$p + C \to \lbrack \Sigma^0K^+  \rbrack + C$ %
small transverse momenta $P^2_T < 0.01 (GeV/c)^2$. %
The state X(1810) is produced only in the region of very small %
$P^2_T$, where it is displayed very clearly and has the parameters %
$M = 1807 \pm 7 MeV , \Gamma  = 62 \pm 19 MeV$. %
The spectrum presented is weighted taking into account the efficiency of the setup; the run is the one with the modified setup. %
}%
\label{fig-5}
\end{figure}
Reaction~(\ref{re-01}) was very clear observed 
in different kinematical regions for $P^2_T$ 
(for all                     $P^2_T$             - Fig.~\ref{fig-3}; 
 in the coherent region      $P^2_T < 0.1 GeV^2$ - Fig.~\ref{fig-4}; 
 in the region of very small $P^2_T$             - Fig.~\ref{fig-5}). 
\begin{figure}
\includegraphics[width=\hsize]{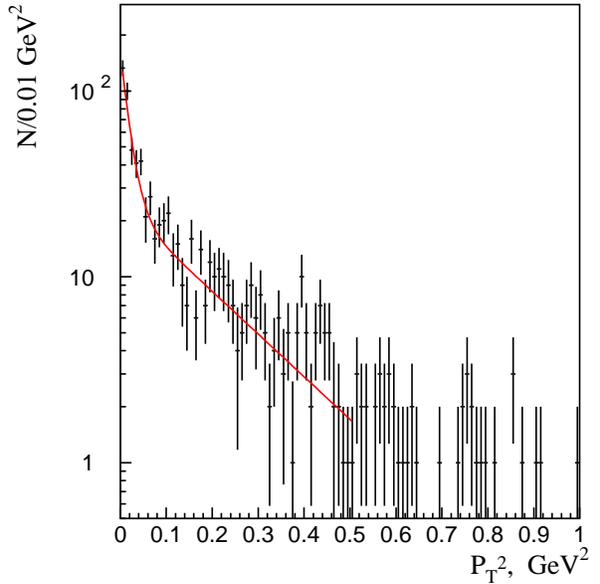}
\caption{%
$dN/dP^2_T$ distribution for the diffractive production reaction %
$p + N \to X(2000) + N$. The distribution is fitted in the form %
$dN/dP^2_T=a_1 \cdot exp(-b_1 \cdot P^2_T) + 
           a_2 \cdot exp(-b_2 \cdot P^2_T)$ %
with the slopes $b_1=63 \pm 10 GeV^2$ (coherent production on carbon nuclei) %
and $b_2=5.8 \pm 0.6 GeV^2$ (production on quasifree nucleons). %
}%
\label{fig-dn_dpt2}
\end{figure}
In Fig.~\ref{fig-dn_dpt2}
 presented the $P^2_T$ distribution 
$dN/dP^2_T$ for the reaction~(\ref{re-01}). 
From this distribution the coherent diffractive production reaction on
carbon nuclei is identified as the narrow peak with the slope
$b_1 \simeq 63 \pm 10 GeV$ 

   The values of cross sections for the reactions under study 
were obtained~\cite{ref_17}: \\
a) X(2000) production in the entire region of $P^2_T$
\begin{eqnarray}
\label{re-04}
\sigma \lbrack p + N \to X(2000) + N \rbrack \cdot
\mathrm{BR}  \lbrack X(2000) \to  \Sigma^0K^+  \rbrack &=&  \nonumber \\
\lbrack 95 \pm 20(stat.) \pm 20(syst.) \rbrack  nbn / nucleon & &
\end{eqnarray}
b) Diffractive coherent production of X(2000) on carbon nuclei
\begin{eqnarray}
\label{re-05}
\sigma \lbrack p + C \to X(2000) + C \rbrack coh. \cdot
 \mathrm{BR} \lbrack X(2000) \to \Sigma^0K^+ \rbrack &=& \nonumber \\
\lbrack 285 \pm 60(stat.) \pm 60(syst.) \rbrack  nbn / nucleus&&
\end{eqnarray}
In the diffractive coherent reaction~(\ref{re-02}) was also observed 
the peak $X(1810) \to \Sigma^0K^+$ with 
$M = 1807 \pm 7 MeV$ and $\Gamma  = 62 \pm 19 MeV$. 
This state is produced only in the region of very small $P^2_T$ 
($< 0.01 \div 0.02 GeV^2$ - see Fig.5). We will not consider possible 
interpretation of this near threshold peak here. This structure needs special 
investigation. 

  Very important property of X(2000) baryon is its large probability of 
the decays with strange particles in the final state.
In the study of the effective mass spectra $M(p\pi^+\pi^-)$ and
$M(\Delta (1232)^{++}\pi^-)$ in the reactions
\begin{eqnarray}
\label{re-06}
p + N (C) & \to & \phantom{D(1} p\pi^+\pi^- \phantom{23)}+ N (C) \\
          & \to & \Delta (1232)^{++}\pi^- + N (C) \nonumber
\end{eqnarray}

    In the same kinematical conditions as in (1) the state X(2000) 
was not observed and lower limits of the corresponding branchings
were obtained
\begin{eqnarray}
\label{re-07}
\mathrm{R1} = \frac
{\mathrm{BR} \lbrack X(2000)^+ \to (\Sigma K)^+ \rbrack} 
{\mathrm{BR} \lbrack X(2000)^+ \to (\Delta \pi)^+ \rbrack}
 > 0.83, \nonumber \\
\mathrm{R2} = \frac
{\mathrm{BR} \lbrack X(2000)^+ \to (\Sigma K)^+ \rbrack} 
{\mathrm{BR} \lbrack X(2000)^+ \to p\pi^+\pi-   \rbrack} 
> 7.8
\end{eqnarray}
(with 95\% C.L.). 

   We also observed the decay $X(2000)^+ \to \Sigma^+K^0$~\cite{ref_18}.
These data are in agreement with main result $X(2000)^+ \to \Sigma^0K^+$ 
in spite of limited statistics for $X(2000)^+ \to \Sigma^+K^0$.
It is reasonable also to mention that in the reaction
$\Sigma^- + N \to  \lbrack \Sigma^-K^+ \rbrack K^- + N$ 
with $P_{\Sigma^-} \simeq 600 GeV$ in the SELEX experiment the state with
$M = 1962 \pm 12 MeV$ and $\Gamma  = 96 \pm 32 MeV$ 
(near the value of the corresponding parameter of X(2000)) was observed 
in the mass spectrum $M(\Sigma^-K^+)$ - see~\cite{ref_02,ref_03}.\\

\section{The experiment with the upgraded SPHINX spectrometer.}
\label{se-3} 

New study of reactions
\begin{eqnarray}
\label{re-00}
p + N (C) \to  \lbrack \Sigma K \rbrack  + N (C) \nonumber
\end{eqnarray}
were carried out with the upgraded SPHINX spectrometer. 
X(2000) baryon was observed in this study on the statistics 
which more than by one order of magnitude exceeded 
the statistics on our previous work~\cite{ref_17}.
 Now we are in the final stage of the data analysis and preparation of 
new results for publication. 
Our preliminary conclusion is that new results are in good agreement 
with previous data. 
\section{The nature of X(2000) baryon and 
         its interpretation as cryptoexotic state with hidden strangeness 
         $\vert X(2000) \rangle = \vert qqqs\bar{s} \rangle$.
}
\label{se-4} 
The main characteristics of X(2000) baryon:\\
4.1\hspace*{0.2cm} The new baryon state X(2000) is produced in 
the SPHINX experiment in gluon-enriched diffractive production reactions 
with Pomeron exchange
\begin{eqnarray}
\label{re-08}
p + N (C) & \to & X(2000)+ + N (C) \\
          &     &\phantom{}\raisebox{.4cm}
{\rotatebox{180}{$\Lsh$}}\Sigma^0K^+; \Sigma^0 \to \Lambda \gamma
                          \nonumber \\
          &     &\phantom{}\raisebox{.4cm}
{\rotatebox{180}{$\Lsh$}}\Sigma^+K^0; \Sigma^+ \to p\pi^0; n\pi^+
                          \nonumber
\end{eqnarray}
in the region of all $P^2_T$, as well as in the coherent diffractive
reaction on carbon nucleus in the region $P^2_T < 0.1 GeV^2$ \\
4.2\hspace*{0.2cm} Due to proton diffractive production of X(2000) baryon
its isotopic spin must be $I = \frac{1}{2}$ and the most probable set 
of quantum numbers is 
$J^P = \frac{1}{2}^+;\frac{3}{2}^-;\frac{5}{2}^+;\frac{7}{2}^-;...$
(Gribov-Morisson selection rule). For $I = \frac{1}{2}$ the ratio of 
the branchings are
\begin{eqnarray}
\label{re-000}
\frac
{\mathrm{BR} \lbrack X(2000)^+ \to \Sigma^0K^+  \rbrack}  
{\mathrm{BR} \lbrack X(2000)^+ \to (\Sigma K)^+ \rbrack}
=\frac{1}{3}; \nonumber \\
\frac
{\mathrm{BR} \lbrack X(2000)^+ \to  \Delta (1232)^{++}\pi^- \rbrack}
{\mathrm{BR} \lbrack X(2000)^+ \to (\Delta (1232)\pi)^+     \rbrack}
=\frac{1}{2}; \nonumber
\end{eqnarray}
4.3\hspace*{0.2cm} The state X(2000) has 
the mass $M = 1986 \pm 6 MeV$ and the width $\Gamma = 98 \pm 21 MeV$. 
It must be stressed that X(2000) has a large one mass with 
a small enough width ($\Gamma / M \simeq 0.05$). 
In the same time all well known ordinary isobars $\vert qqq \rangle$ 
with close masses have large widths $\Gamma  > 300 \div 500 MeV$ 
(see~\cite{ref_23} and Table~\ref{tab-1}).\\
\begin{table*}
\begin{minipage}[h]{\hsize}
\centerline{
\begin{tabular}{|l|c|c|c|c|c|c|c|}
\hline\noalign{\smallskip}
$N^*$          &                          &\multicolumn{6}{|c|}{}   \\ 
$J^P(status)$  &  $\Gamma/(MeV)$            &\multicolumn{6}{|c|}{%
                 Probability ratio for different decay channels  }  \\
&&
     $\frac{N^* \to N     \pi}{N^* \to tot}$         &
     $\frac{N^* \to \Delta\pi}{N^* \to tot}$         &
     $\frac{N^* \to N\pi  \pi}{N^* \to tot}$         &
     $\frac{N^* \to \Sigma  K}{N^* \to tot}$         &
     $\frac{N^* \to \Sigma  K}{N^* \to \Delta\pi}$   &
     $\frac{N^* \to \Sigma  K}{N^* \to \Delta\pi}$                  \\
\noalign{\smallskip}\hline\noalign{\smallskip}
N(1900)                    & $498 \pm 78 $                        &
0.26                       & \multicolumn{2}{|c|}{0.4}            &
\rule[0mm]{1cm}{0.2mm}     &\rule[0mm]{1cm}{0.2mm}                &
\rule[0mm]{1cm}{0.2mm}                                              \\
$3/2^+(*)$                 &  \multicolumn{7}{|c|}{}                \\
\noalign{\smallskip}\hline\noalign{\smallskip}
N(1990)                    & $200\div 500 $                       &
0.6                        &  \multicolumn{2}{|c|}{$0.3\div 0.9 $}&
$(2  \div 60)\cdot 10^{-3}$&  \multicolumn{2}{|c|}{$<0.1$}          \\
$7/2^+(*)$                 &  \multicolumn{7}{|c|}{}                \\
\noalign{\smallskip}\hline\noalign{\smallskip}
N(2000)                    & $490\pm  310 $                       &
0.6                        & $0.15\div 0.20$                      &
$0.6\div 0.7$              &
$(1  \div  4)\cdot 10^{-2}$&
$(6  \div 25)\cdot 10^{-2}$&
$(1  \div  3)\cdot 10^{-2}$                                         \\
\noalign{\smallskip}\hline\noalign{\smallskip}
N(2080)                    & $200\pm  600 $                       &
$0.13\div 0.16$            & $0.25\div 0.30$                      &
$0.5\div 0.6$              &
$(1.5\div 40)\cdot 10^{-3}$&
$(6  \div 40)\cdot 10^{-3}$&
$(3  \div 20)\cdot 10^{-3}$                                         \\
$3/2^-(**)$                &  \multicolumn{7}{|c|}{}                \\
\noalign{\smallskip}\hline\noalign{\smallskip}
N(2090)                    & $414 \pm 185 $                       &
$0.10 \div 0.15$           &\rule[0mm]{1cm}{0.2mm}                &
\rule[0mm]{1cm}{0.2mm}     &\rule[0mm]{1cm}{0.2mm}                &
 \multicolumn{2}{|c|}{}                                             \\
$1/2^-(*)$                 & $350 \pm 100 $                       &  
\multicolumn{6}{|c|}{}                                              \\
                           & $ 95 \pm  30 $                       &
\multicolumn{6}{|c|}{}                                              \\
\noalign{\smallskip}\hline\noalign{\smallskip}
N(2100)                    & $113 \pm  40 $                       &
$0.10 \div 0.15$           & 0.4                                  &
\rule[0mm]{1cm}{0.2mm}     &\rule[0mm]{1cm}{0.2mm}                &
 \multicolumn{2}{|c|}{}                                             \\
$1/2^+(*)$                 & $260 \pm  10 $                       &  
\multicolumn{6}{|c|}{}                                              \\
                           & $200 \pm  30 $                       &
\multicolumn{6}{|c|}{}                                              \\
\noalign{\smallskip}\hline\noalign{\smallskip}
N(2190)                    & $450\pm  100 $                       &
$0.1 \div 0.2 $            &  \multicolumn{2}{|c|}{$\gtrsim 0.4$} &
$(1.5\div  3)\cdot 10^{-3}$&
  \multicolumn{2}{|c|}{$(3  \div  6)\cdot 10^{-3}$}                 \\
$7/2^-(****)$\footnote{**** - well established isobar state; %
* or ** - only some weak evidence of their existence are obtained}
              &  \multicolumn{7}{|c|}{}                \\
\noalign{\smallskip}\hline\noalign{\smallskip}
\end{tabular}
}
\end{minipage}
\caption{%
Properties of massive isobars $N^*= \vert qqq \rangle (q=u, d-quarks)$.%
}
\label{tab-1}
\end{table*}
4.4\hspace*{0.2cm} The X(2000) state has been experimentally shown
to decay mainly via channels involving the emission of strange particles
%
\begin{eqnarray}
\label{re-09}
\mathrm{R}=\frac
{\mathrm{BR} \lbrack X(2000) \to \Sigma K \rbrack} 
{\mathrm{BR} \lbrack X(2000) \to p\pi^+\pi^-, \Delta (1232)\pi \rbrack}
\gtrsim 1
\end{eqnarray}
while the decays $X(2000) \to p\pi^+\pi-, \Delta (1232)\pi$ are strongly 
suppressed (by two order of magnitude) - see~\cite{ref_23} and 
Table~\ref{tab-1}.
Thus the X(2000) baryon exhibit anomalous dynamical properties 
that can not be explained if it is interpreted 
as ordinary $\vert qqq\rangle$ baryon. 
But all these anomalies can be readily explained if this state
is assumed to be a cryptoexotic pentaquark baryon %
with hidden strangeness $\vert qqqs\bar{s}\rangle$.
It must be specified more exactly that when we mention 
ordinary baryon having the structure $\vert qqq \rangle$,         or 
exotic baryon                        $\vert qqqs\bar{s} \rangle$, or 
                                     hybrid $\vert qqqg \rangle$, 
we actually mean only those hadronic components that determine their principal
characteristics (quantum numbers, main dynamical properties). 
They are called valence quarks and gluons.

Any hadron also contains a quark-gluon "sea" of virtual gluons and 
quark-antiquark pairs emitted and absorbed by the valence structure elements. 
The quark-gluon "sea" determines many hadron properties 
(such as, for example, the spacial distribution of electric charges and
magnetic moments inside the particles). 
In the region of large distances compared with the dimensions of hadron
(i.e. in the region of small enough $P^2_T$ ) they may behave 
like system composed of valence quarks. 
When studies of phenomena of relatively large transverse momentum
(small distances) are performed (i.e. when the structure of hadron is 
investigated) manifestations of quark-gluon "sea" are arise. 
This picture is qualitatively consistent with quantum-chromodynamic ideas. 
Therefore when we speak of the complex quark or 
quark-gluon structure of cryptoexotic baryons, 
we actually mean the valence composition of these particles 
which determines their quantum numbers and their dynamical properties 
that are displayed at large distances 
(decay width, branching ratios of decay channels, 
mechanisms of particle production and so on).

   The study of anomalous dynamical properties of X(2000) baryon and 
its interpretation as possible pentaquark baryon with hidden strangeness 
is performed with help of OZI selection rule, or the selection rule 
relative to continuous quark lines~\cite{ref_24}. 
In accordance with OZI rule in the allowed processes the quark lines 
of the respective diagrams depicting one or another process are continuous. 
Processes involving annihilation or creation of quark-antiquark pairs
entering into composition of the same hadron are strongly suppressed. 
The OZI rule suppression factors in different reactions are in the region of
$f \sim 5 \cdot 10^{-2} \div 5 \cdot 10^{-3}$~\cite{ref_26}.
%
\begin{figure}
\includegraphics{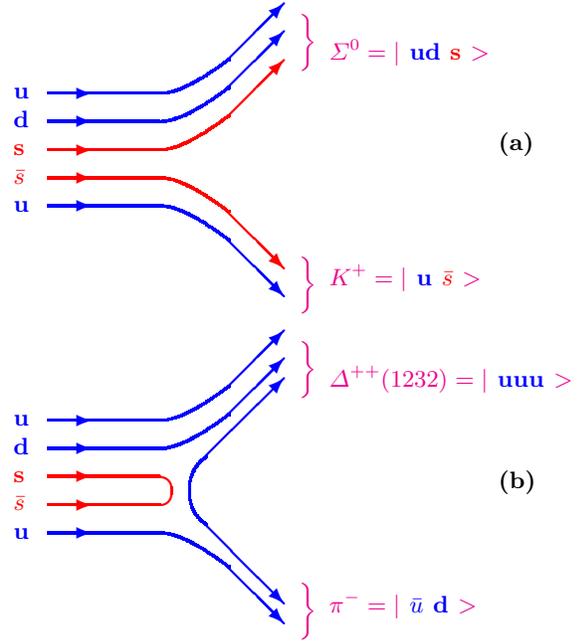}
\caption{%
Diagrams for decays of exotic baryons with hidden strangeness %
$\vert qqqs\bar{s} \rangle$ : %
(a) allowed by OZI rule (decays with emission of strange particles) %
(b) forbidden by OZI rule (decays with emission of non-strange particles)%
}%
\label{fig-Xdecay}
\end{figure}
As it is seen from Fig.~\ref{fig-Xdecay} the decays of 
$\vert qqqs\bar{s} \rangle$ pentaquark with the emission of strange particles 
is OZI allowed, but the decay 
$\vert qqqs\bar{s} \rangle \to \Delta (1232)\pi, p\pi^+\pi^-$ - are
OZI forbidden and thus are strongly suppressed. 
This makes it possible to explain, why in the case of exotic baryons
with hidden strangeness, the decays via channels with strange particles 
provide to be large and even to be dominant.

Moreover, in the case of allowed decays 
$\vert qqqs\bar{s} \rangle \to \Sigma K$ 
owing to relatively large mass of the particles, 
the kinetic energy released in the decay turns out to be significantly
lower than in the decay $\vert qqqs\bar{s} \rangle \to \Delta (1232)\pi$. 
This kinematical factor, as well as more complex internal 
structure of the exotic baryons may significantly reduced 
its total decay width. 
Thus, anomalous properties of X(2000) baryon make it a strong candidate 
for cryptoexotic pentaquark baryon with hidden strangeness.

I am clearly understand that observation of even strong candidate 
for for cryptoexotics by some indirect dynamical properties 
do not allow one to claim unambiguously that this state is really 
a new cryptoexotic form of hadronic matter. 
May be it still possible to find another interpretation 
of these anomalous properties. 
This is general problem not only for the X(2000) baryon, 
but for all known meson and baryon candidates for cryptoexotics 
(for example, for the candidate to hybrid meson $\pi(1800)$~\cite{ref_25}). 
In all these cases we need additional studies and new arguments.

Thus, let us to try to direct the way for future study of X(2000) baryon 
to produce new information for the nature of this hadron. 
Certainly, we must determine its quantum numbers and to investigate 
another decay channels. 
But it seems that the decisive arguments will be obtained 
in the studies of different mechanisms for X(2000) production.

\begin{figure}
\vspace*{-2cm}
\includegraphics{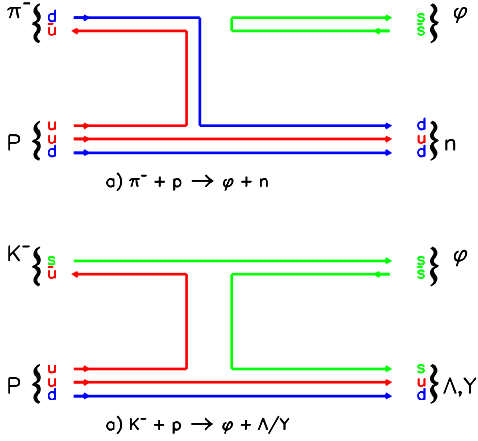}
\vspace*{-2cm}
\caption{%
Diagrams for the reaction $K/\pi^{-} + p \to \phi + \Lambda , Y/n$
}
\label{fig-phi}
\end{figure}

Let us remain first of all some data for the production
of well known $\phi$ - meson with hidden strangeness 
($\vert \phi \rangle \approx \vert s\bar{s} \rangle$) 
in the pion and kaon beams in reactions
\begin{eqnarray}
\label{re-10}
\pi^- + p \to \phi + n
\end{eqnarray}
and
\begin{eqnarray}
\label{re-11}
K^- + p \to \phi + Y
\end{eqnarray}
%
As seen from the diagrams in Fig.~\ref{fig-phi} the $\phi$ production 
in (10) is OZI forbidden and strongly suppressed and in (11) is OZI allowed. 
From the experimental data of~\cite{ref_26} for $M(K^+K^-)$ mass spectra 
in reaction~(\ref{re-10}) and~(\ref{re-11}) at $P_{\pi;K} = 32.5 GeV$ 
\begin{figure}
\includegraphics[width=\hsize]{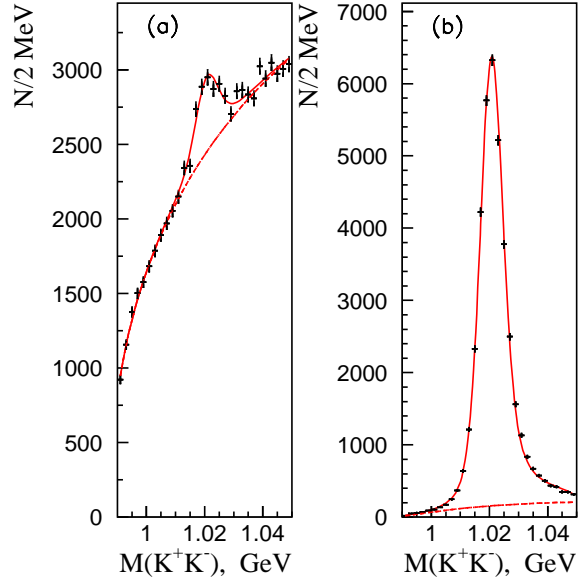}
\caption{%
The data for $M(K^+K^-)$ and separation of $\phi$-meson in : %
(a) OZI - forbidden reaction $\pi^- + p \to \phi + n$;
(b) OZI - allowed reaction $    K^- + p \to \phi + Y$;
}
\label{fig-9ab}
\end{figure}
\begin{figure}
\vspace*{-2cm}
\includegraphics[width=\hsize]{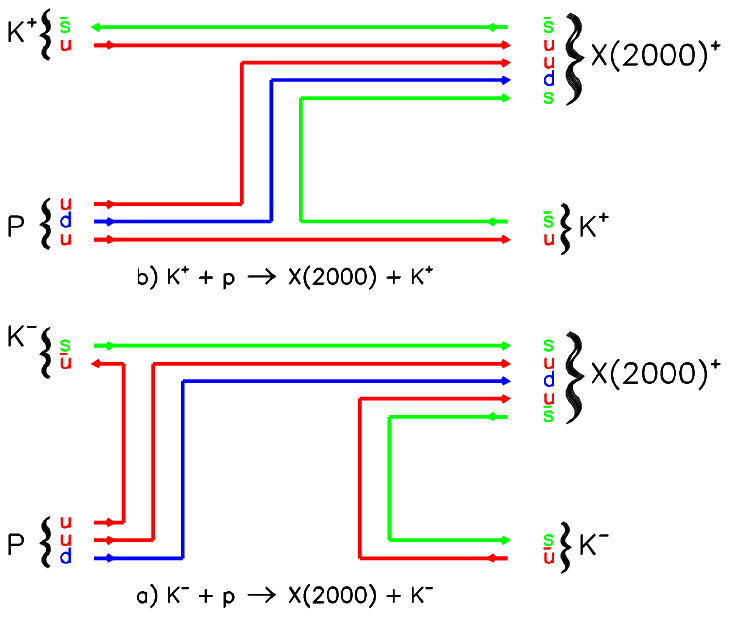}
\vspace*{-1.5cm}
\caption{%
Reaction $K^{\pm} + p \to X(2000) + K^{\pm}$ %
for $\vert X \rangle = \vert uuds \bar{s}\rangle$ %
}
\label{fig-kpm-x2000}
\end{figure}
(Fig.~\ref{fig-9ab} and Fig.10), it is seen that ratio of cross sections is
\begin{eqnarray}
\label{re-12}
\frac{\sigma(\pi^- + p \to \phi + n)}
     {\sigma(   K- + p \to \phi + Y)}
\sim 10^{-2}
\end{eqnarray}
(the cross section $\sigma(\pi^- + p \to \phi + n)
\vert_{P_{\pi^-} =32.5GeV} = 11.5 \pm 3.3 nbn$)

\begin{figure}
\vspace*{-2cm}
\includegraphics{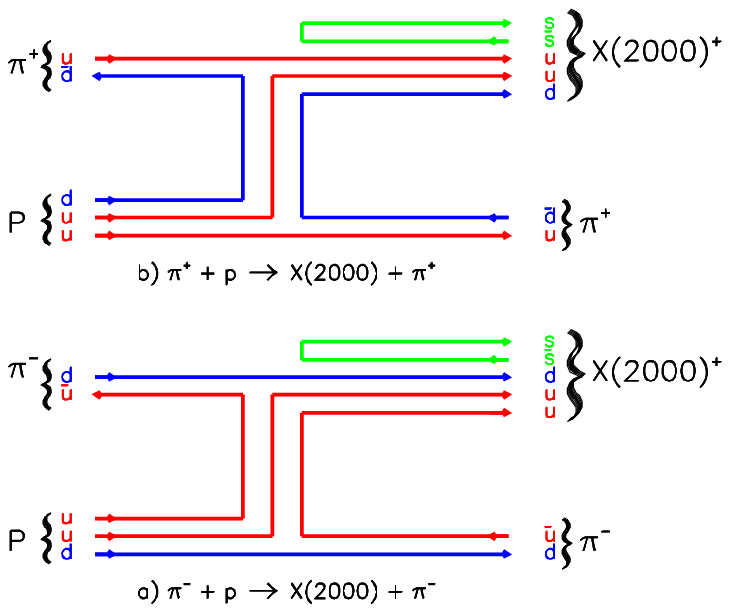}
\vspace*{-2cm}
\caption{%
Reaction $\pi^{\pm} + p \to X(2000) + pi^{\pm}$ %
for $\vert X \rangle = \vert uuds \bar{s}\rangle$ %
}
\label{fig-pipm-x2000}
\end{figure}

We propose to use the same technic of studying 
the hidden strangeness in the valent quark system of hadrons
for X(2000) baryon in the reactions with baryon exchange (backward scattering)
\begin{eqnarray}
\label{re-13}
\pi^{\pm} + p \to X(2000)+ + \pi^{\pm}
\end{eqnarray}
and
\begin{eqnarray}
\label{re-14}
K^{\pm} + p \to X(2000)^+ + K^{\pm}
\end{eqnarray}
If X(2000) baryon is really pentaquark state with hidden strangeness 
($\vert X(2000)\rangle = \vert qqqs\bar{s}\rangle$), the reaction~(\ref{re-13})
are OZI forbidden and strongly suppressed and 
the reaction~(\ref{re-14}) is OZI allowed 
(see diagrams in Fig.~\ref{fig-pipm-x2000} and Fig.~\ref{fig-kpm-x2000}).
Thus, in this case it is possible to expect that the ratio of cross sections is
\begin{eqnarray}
\label{re-15}
\frac{\sigma(\pi^{\pm} + p \to X(2000)^+ + \pi^{\pm})}
     {\sigma(  K^{\pm} + p \to X(2000)^+ +   K^{\pm})} \ll 1
\end{eqnarray}
%
May be the ratio~(\ref{re-15}) is not too small as for $\phi$ - meson, 
but significantly enough to obtain the definite conclusion 
about the hidden strangeness of X(2000) baryon.
Such experiment is of crucial importance for pentaquark
interpretation of X(2000) baryon.

   The experimental studies of reaction~(\ref{re-13}) and~(\ref{re-14})
with baryon exchange can be performed with the new OKA experiment on 
the IHEP kaon separated beam with momenta 12.5 and 
17 GeV~\cite{ref_27,ref_28}. 
This momentum region seems to be optimal for studying baryon exchange
processes as a compromise between 
the drop of cross section of these reactions with energy and 
the increasing of the efficiency for their registration.
\begin{acknowledgement}
{\Large \bf{   Acknowledgements}}\\
It is a pleasure to express my gratitude to 
L.B.~Okun for stimulated discussion and
D.I.~Patalakha for his help in preparation of this paper.
This work was partly supported by Russian
Foundation for Basic Researches (grants 05-02-16924-a).
\end{acknowledgement}

\end{document}